\documentclass[a4paper]{jpconf}
\usepackage{graphicx}
\begin{document}
\title{Heavy-flavour production in proton--proton collisions with the ALICE experiment}

\author{László Gyulai (for the ALICE Collaboration)}

\address{Wigner Research Centre for Physics, Budapest, Hungary}

\ead{laszlo.gyulai@cern.ch}

\begin{abstract}
The production of heavy-flavour hadrons in high-energy hadronic collisions is a unique source of information on various aspects of quantum chromodynamics (QCD). Production of heavy-flavour hadrons in proton--proton collisions allows the test of perturbative QCD models, while the comparison of mesons and baryons with heavy-flavour quarks can differentiate between fragmentation scenarios. Multiplicity-dependent measurements allow for the understanding of semi-hard vacuum QCD effects, as well as to study the coalescence mechanisms of heavy-flavour quarks with light and strange quarks. Recent results from the ALICE experiment in proton--proton collisions on the production of D mesons and leptons from the decay of heavy-flavour hadrons, as well as charmed baryons, are presented in this contribution. Furthermore, the multiplicity dependence of self-normalised heavy-flavour electron yields, as well as that of strange to non-strange D-meson and charmed baryon-to-meson ratios are also shown.
\end{abstract}

\section{Introduction}

Heavy-flavour quarks (charm and beauty) are produced in the initial stage of heavy-ion collisions via partonic hard scatterings. Therefore they can be used to study the initial hard processes and the properties of the hot and cold nuclear medium present in high-energy heavy-ion collisions. In proton--proton (pp) collisions, heavy-flavour particles can be primarily used for testing calculations based on perturbative quantum chromodynamics (pQCD), study flavour-dependent fragmentation, and to set a baseline for nuclear modification in heavy-ion collisions. Besides this, comparing the production yield of different heavy-flavour particle species (baryons to mesons, strange to non-strange heavy-flavour hadrons) provides the opportunity to study the hadronization processes. Unexpected was the discovery of a collectivity-like phenomenon in small collision systems (pp, p--Pb) with high final-state multiplicity \cite{Khachatryan:2016txc}. Collectivity was observed prior to that in heavy-ion collisions, where it is attributed to the presence of the strongly-interacting quark--gluon plasma (QGP) \cite{Adcox:2004mh}. It is unlikely, however, that the QGP is produced in a substantial volume in pp collisions, due to insufficient energy density. To better understand the origin of the collectivity-like phenomenon in small collision systems, multiplicity-dependent studies of pp collisions are performed. Recent results show that the collectivity-like phenomenon can be explained with vacuum QCD effects on the soft-hard boundary, such as Multiple Parton Interactions (MPI) \cite{Ortiz:2013yxa}.

The ALICE experiment \cite{Aamodt:2008zz} provides a great opportunity for studying heavy-flavour hadrons due to its high-precision tracking system. The Inner Tracking System is a set of detectors based on silicon technology, which can reconstruct decay vertices of heavy-flavour hadrons with precision down to 100 $\mu$m. The Time Projection Chamber, along with the Time of Flight Detector, provide charged-particle tracking and identification, while high-$p_{\rm T}$ electrons are further identified with Electromagnetic Calorimeter. These detectors cover the midrapidity region ($|\eta|<0.5$). Muons, on the other hand, are detected at forward rapidity ($2.5<\eta<4$) by the muon spectrometer. In the ALICE experiment, heavy-flavour hadrons are detected via two types of decay channels: hadronic decays and semi-leptonic decays. In the first case a full kinematical reconstruction is performed from decay products. The invariant mass distribution of the hadron candidates is then fitted to discard the combinatorial background. The hadronic decay channels of charmed hadrons in the current analyses are ${\rm D^0\rightarrow K^-\pi^+}$, ${\rm D_{s}^+\rightarrow K^-K^+\pi^+}$, ${\rm \Lambda_{c}^+\rightarrow pK^-\pi^+}$, ${\rm \Sigma_{c}^0\rightarrow \Lambda_{c}^+\pi^-}$, ${\rm \Sigma_{c}^{++}\rightarrow \Lambda_{c}^+\pi^+}$, ${\rm \Xi_{c}^0\rightarrow \Xi_{c}^-\pi^+}$, ${\rm \Xi_{c}^+\rightarrow \Xi_{c}^-\pi^+\pi^+}$. The production of beauty quarks in ALICE is accessed via the non-prompt measurements of D mesons originating from the decay of beauty hadrons. In case of the investigation of semi-leptonic decay channels, electrons from charm and beauty quarks are reconstructed in the central barrel, and their contributions are statistically separated on the basis of the impact parameter distribution of decay electrons, while muons from charm and beauty quarks are reconstructed in the forward muon spectrometer.

\section{Production of heavy-flavour hadrons in pp collisions}

\begin{figure}[ht!]
\begin{minipage}{0.45\textwidth}
\includegraphics[width=\textwidth]{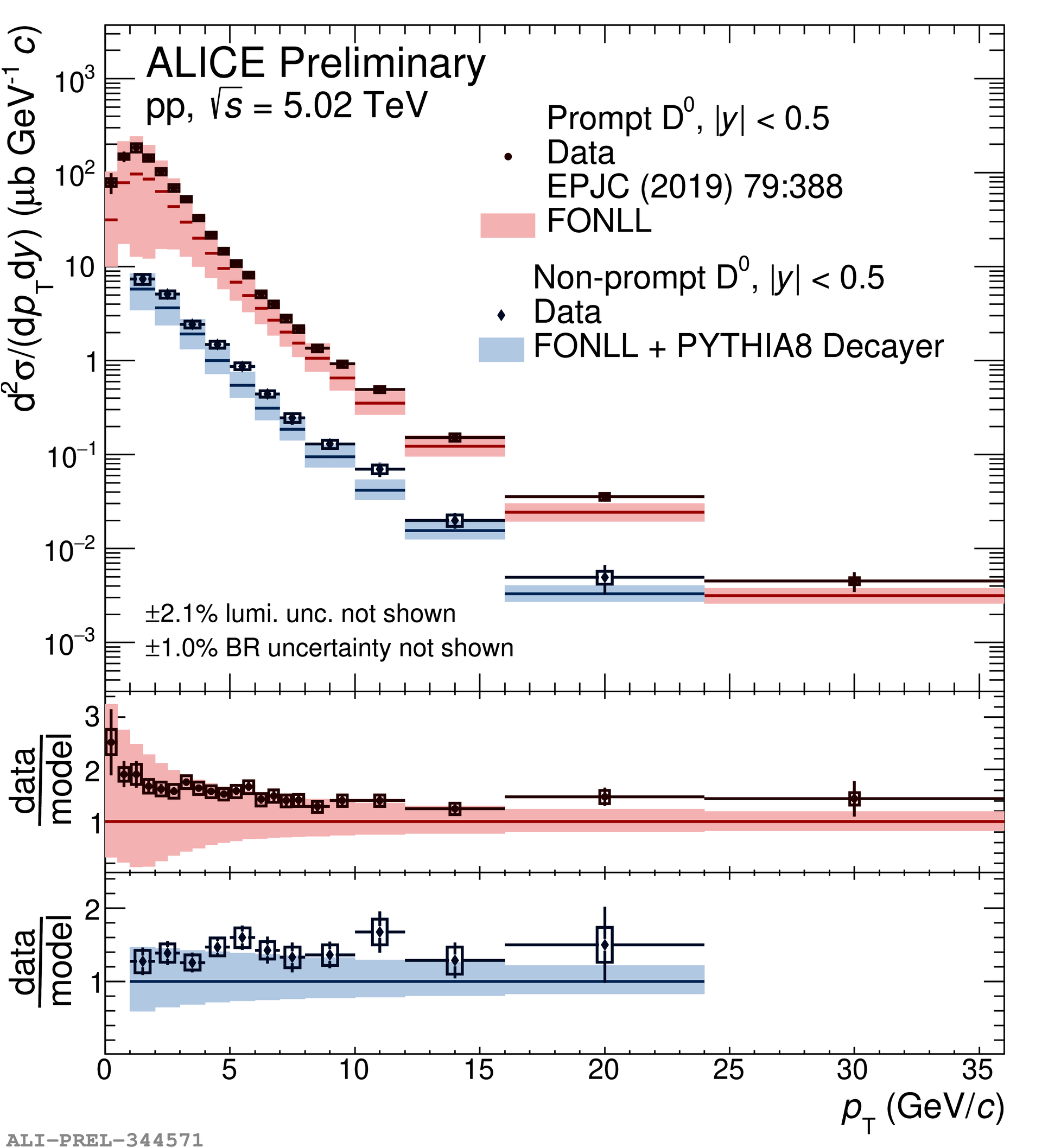}
\caption{\label{D0cross}Production cross section of prompt and non-prompt $\rm{D^0}$ mesons compared with the FONLL calculations.}
\end{minipage}\hspace{0.1\textwidth}%
\begin{minipage}{0.45\textwidth}
\includegraphics[width=\textwidth]{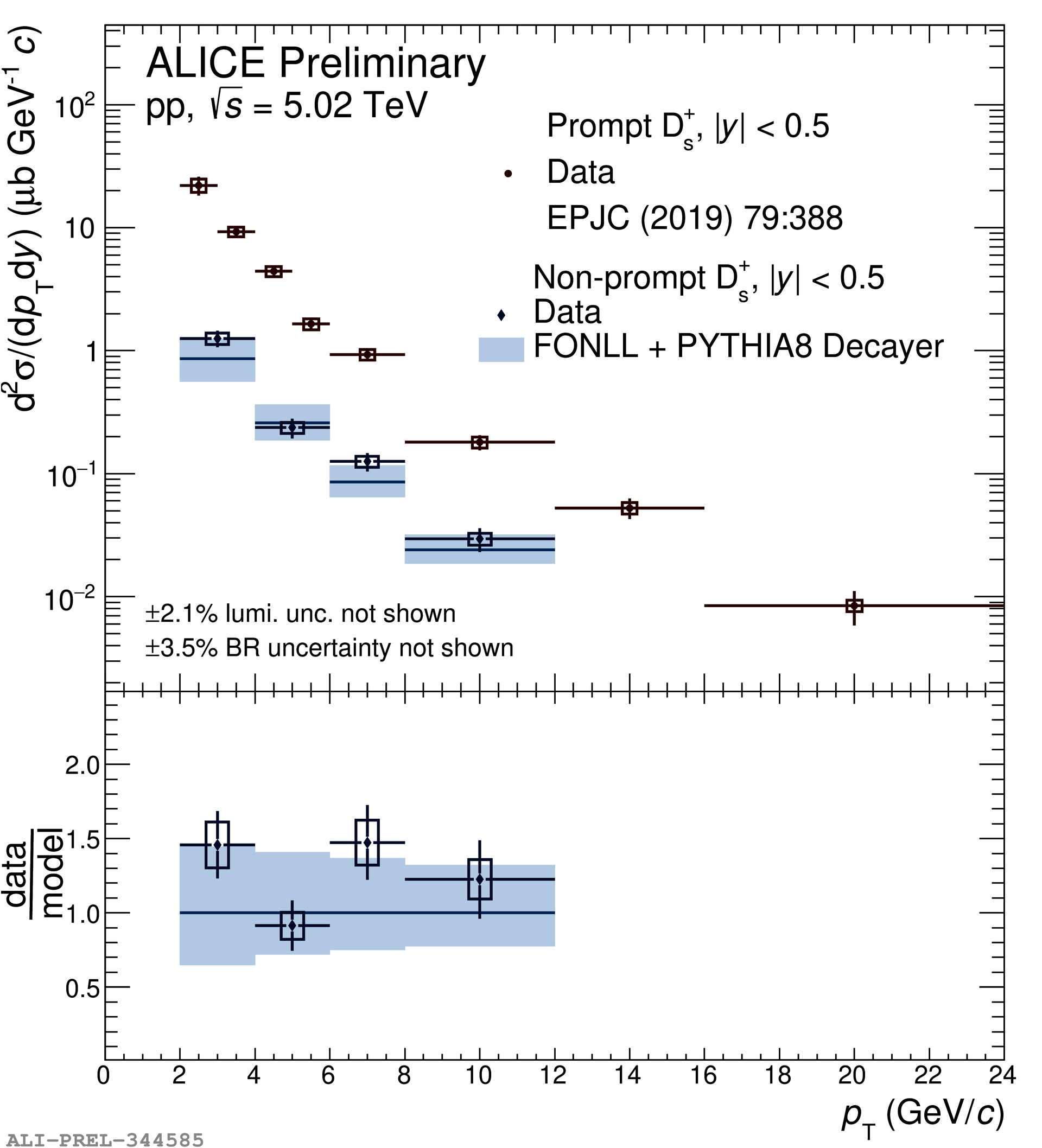}
\caption{\label{D+cross}Production cross section of prompt and non-prompt $\rm{D_s^+}$ mesons compared with the FONLL calculations.}
\end{minipage} 
\end{figure}

The production of prompt and non-prompt D mesons in pp collisions at $\sqrt{s}=5.02$ TeV was measured with ALICE detectors. In Fig.~\ref{D0cross} the production cross section of prompt ${\rm D^0}$ mesons \cite{Acharya:2019mgn}, as well as preliminary results for non-prompt ${\rm D^0}$ mesons are compared to FONLL calculations \cite{Cacciari:1998it}. The perturbative QCD model describes well the results, with a trend of data being closer to the upper edge of the FONLL uncertainties. The production cross section of non-prompt strange D mesons is also well described by the FONLL model (Fig.~\ref{D+cross}).

\begin{figure}[ht!]
\begin{minipage}{0.45\textwidth}
\includegraphics[width=\textwidth]{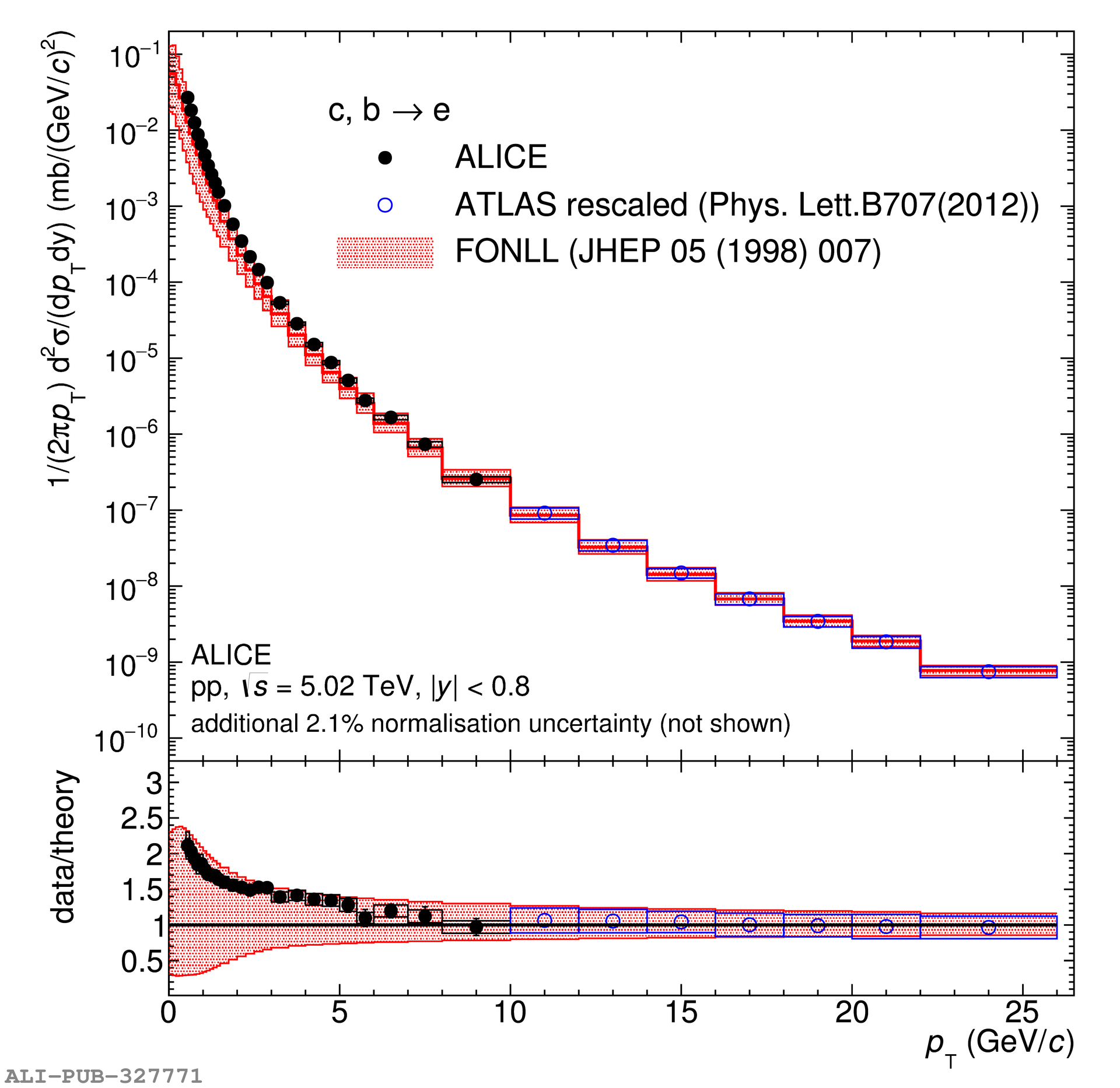}
\caption{\label{ecross}Production cross section of electrons from semileptonic heavy-flavour hadron decays compared with the FONLL calculations.}
\end{minipage}\hspace{0.1\textwidth}%
\begin{minipage}{0.45\textwidth}
\includegraphics[width=\textwidth]{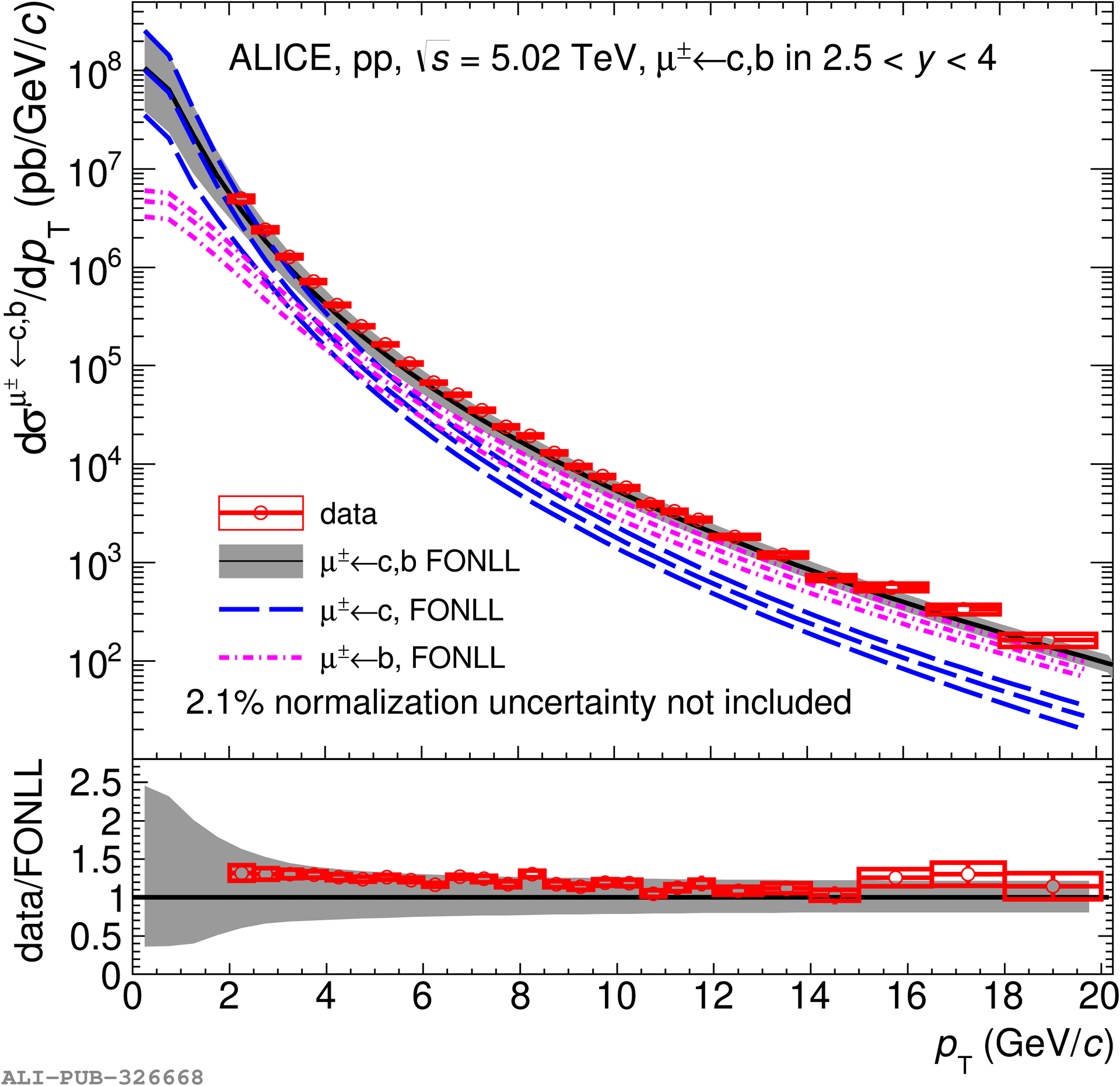}
\caption{\label{mucross}Production cross section of muons from semileptonic heavy-flavour hadron decays compared with the FONLL calculations.}
\end{minipage} 
\end{figure}

Figure \ref{ecross} shows the production cross section of electrons from heavy-flavour hadron decays measured in ALICE at midrapidity up to $p_{\rm T}$=10 GeV/$c$ \cite{Acharya:2019mom} and the rescaled ATLAS measurement \cite{Aad:2011rr}. The results are compared with the FONLL predictions. Similar to the behaviour observed for D mesons, in electron measurements a trend of data being closer to the upper edge of the FONLL uncertainty is also seen at low and intermediate $p_{\rm T}$. At high $p_{\rm T}$, data are closer to the central values of the FONLL predictions. Measurements of muons from heavy-flavour hadron decays are shown in Fig.~\ref{mucross}. FONLL predicts more muons from charm hadron decays at low $p_{\rm T}$, while at high transverse momenta muons from beauty hadron decays are dominant. The production of muons from heavy-flavour decays, measured at forward rapidity, is also well described by the FONLL calculations. High-precision ALICE measurements on heavy-flavour production already provide strong restrictions on theoretical calculations.

\section{Fragmentation of heavy-flavour quarks: mesons and baryons}

\begin{figure}[ht!]
\begin{minipage}{0.45\textwidth}
\includegraphics[width=\textwidth]{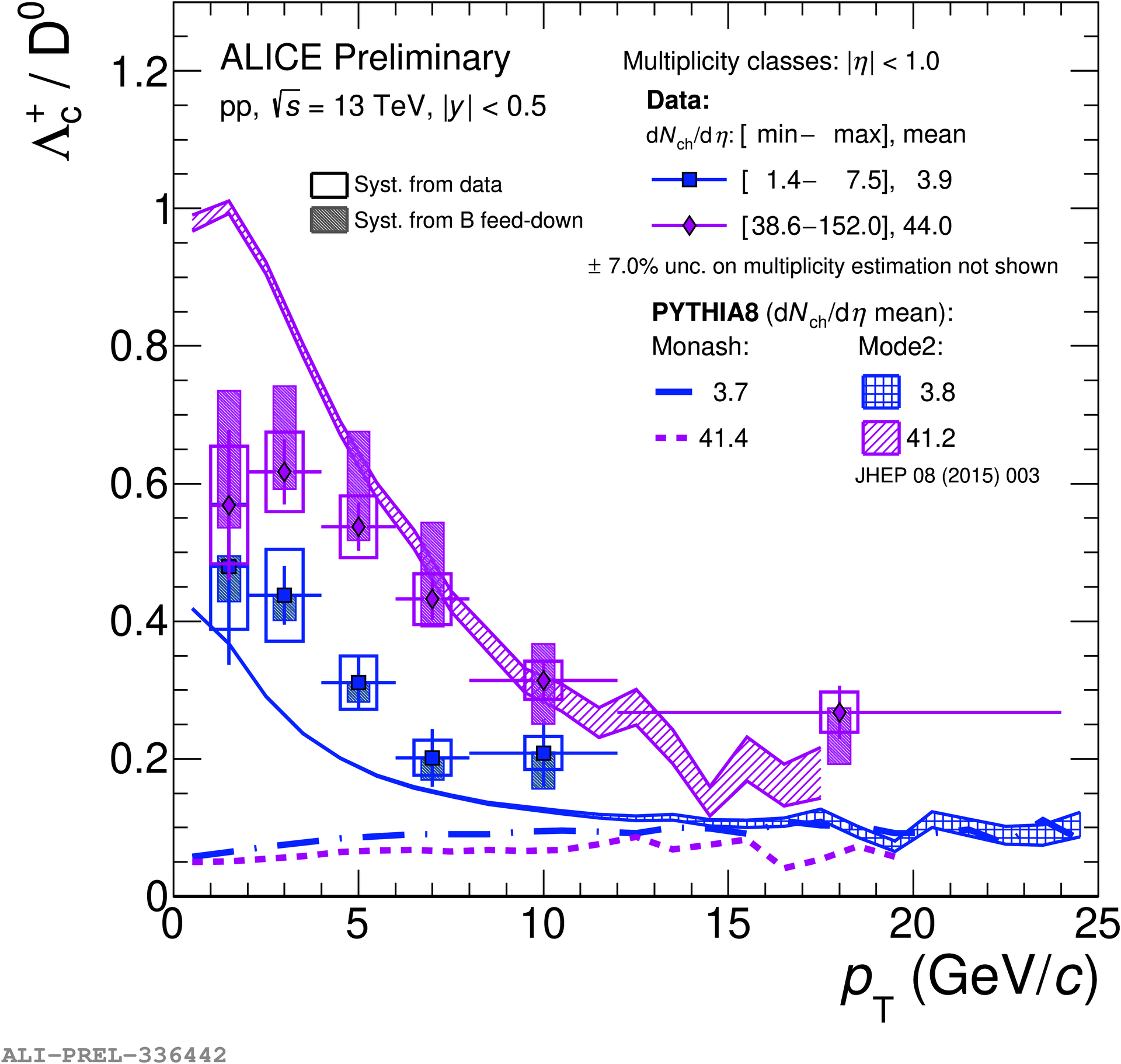}
\caption{\label{LoverD}Ratio of ${\rm \Lambda_c}$ to $\rm{D^0}$ at low and high multiplicity compared to the PYTHIA8 simulations with Monash and Mode2 tunes.}
\end{minipage}\hspace{0.1\textwidth}%
\begin{minipage}{0.45\textwidth}
\includegraphics[width=\textwidth]{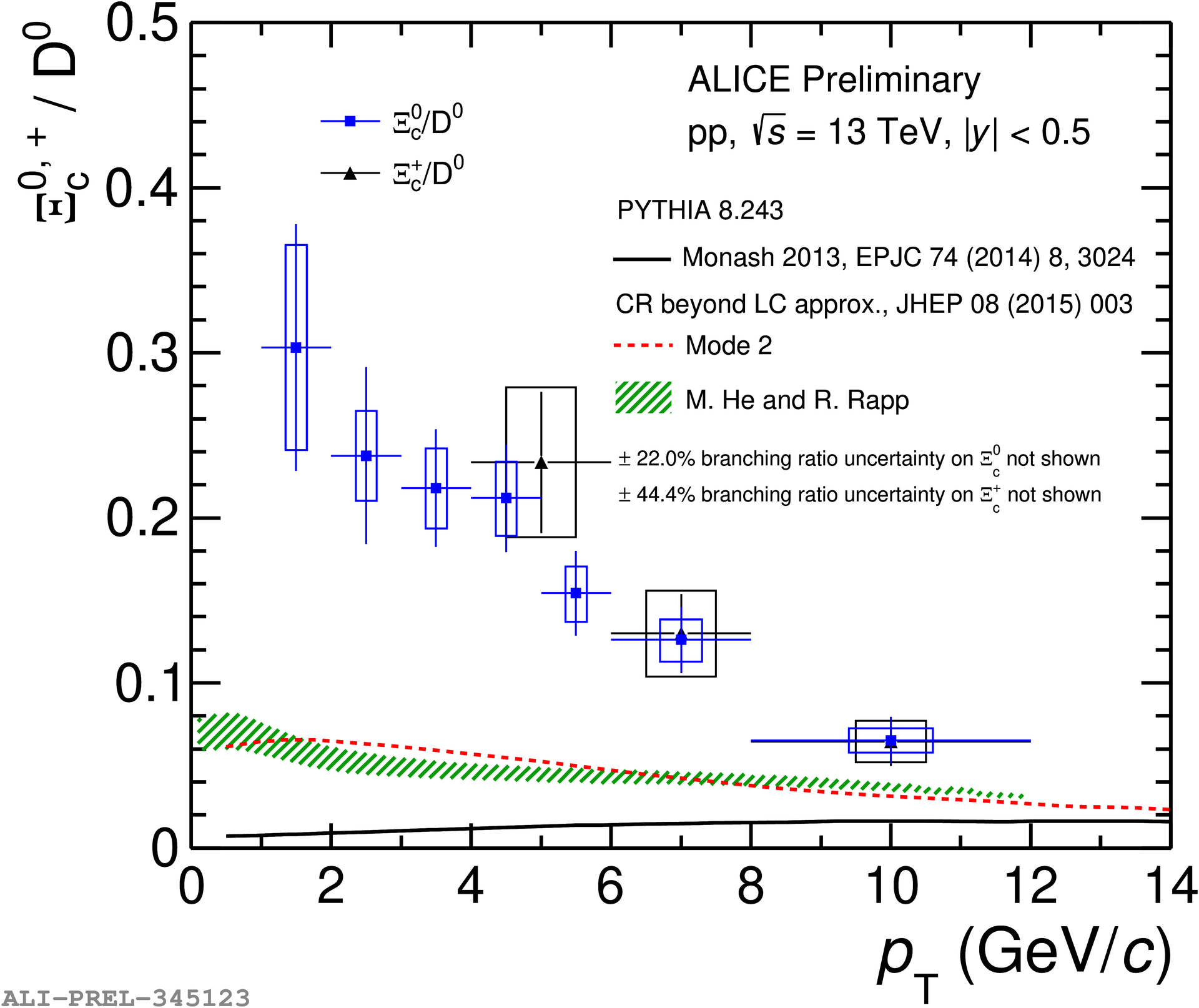}
\caption{\label{XoverD}Ratio of ${\rm \Xi_c^{0,+}}$ to $\rm{D^0}$ compared to the PYTHIA8 simulations with Monash and Mode2 tunes, as well as to the SHM-based model.}
\end{minipage} 
\end{figure}

Differences in fragmentation of c and b quarks into baryons and mesons can be studied by calculating the ratios of different particle species. Figure \ref{LoverD} shows the production ratio of ${\rm \Lambda_c^+}$ baryons to ${\rm D^0}$ mesons. An enhancement is observed at low $p_{\rm T}$, which is not described by the PYTHIA8 Monash tune \cite{Sjostrand:2014zea, Skands:2014pea}. However, the PYTHIA8 Mode2 with colour reconnection using string formation beyond leading colour approximation \cite{Christiansen:2015yqa} provides a qualitatively good description of data.

Figure \ref{XoverD} shows the ratio of ${\rm \Xi_c^0}$ and ${\rm \Xi_c^+}$ baryons, containing a strange quark, to ${\rm D^0}$ mesons. The PYTHIA8 Monash tune does not predict the enhancement of data at low $p_{\rm T}$, while PYTHIA8 Mode2 significantly underestimates it. Another model proposed to describe the hadronization of charmed baryons with the statistical hadronization model (SHM) by employing an augmented set of charm-baryon states beyond the current listings of the particle data group \cite{He:2019tik} also fails in reproducing the relative enhancement of strange charmed-baryons at low transverse momenta.

\begin{figure}[ht!]
\begin{minipage}{0.48\textwidth}
\includegraphics[width=\textwidth]{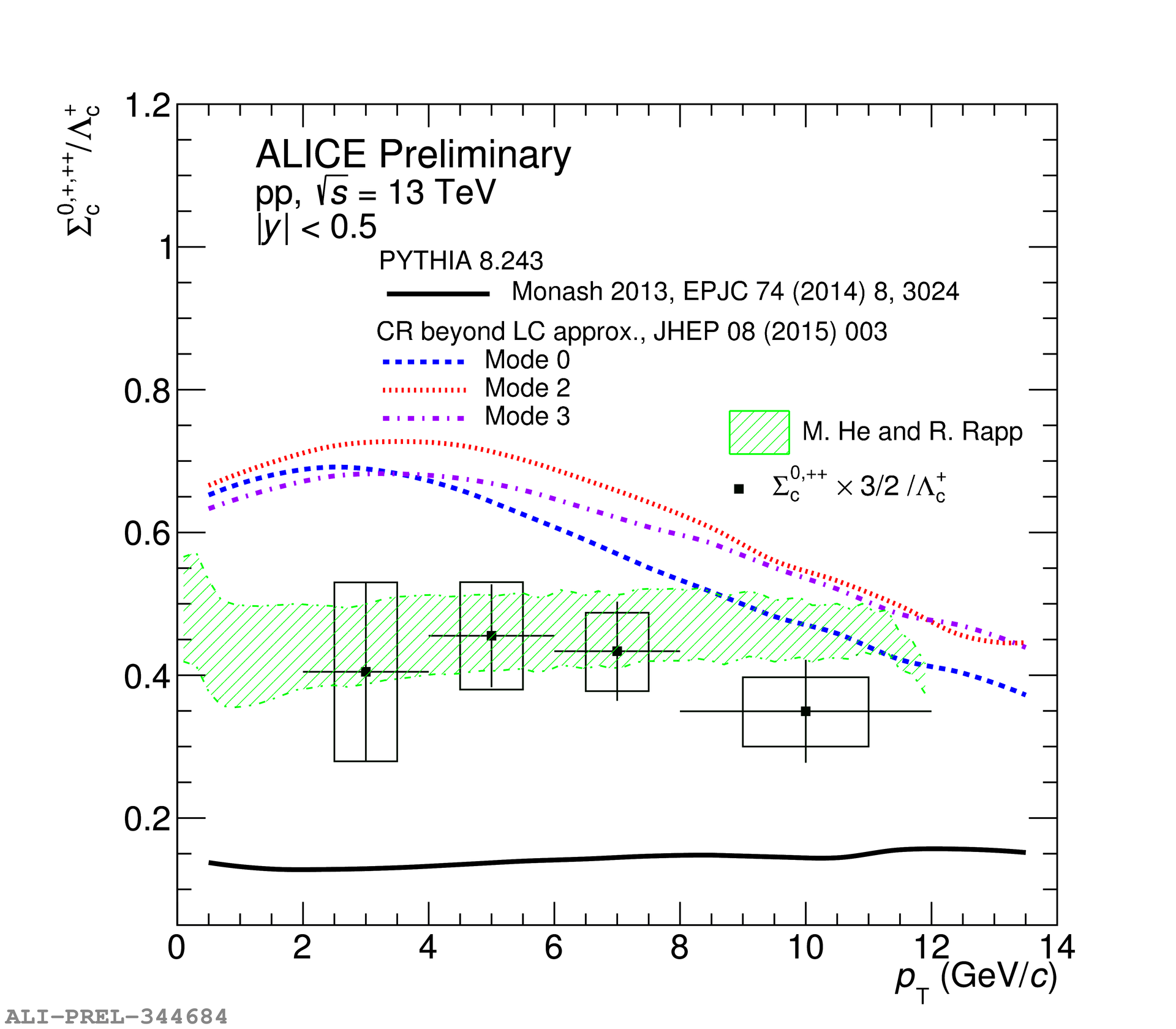}
\caption{\label{SoverL}Ratio of ${\rm \Sigma_c}$ and ${\rm \Lambda_c}$ baryons compared with PYTHIA8 predictions with different tunes, as well as with SHM-based model.}
\end{minipage}\hspace{0.04\textwidth}%
\begin{minipage}{0.48\textwidth}
\includegraphics[width=\textwidth]{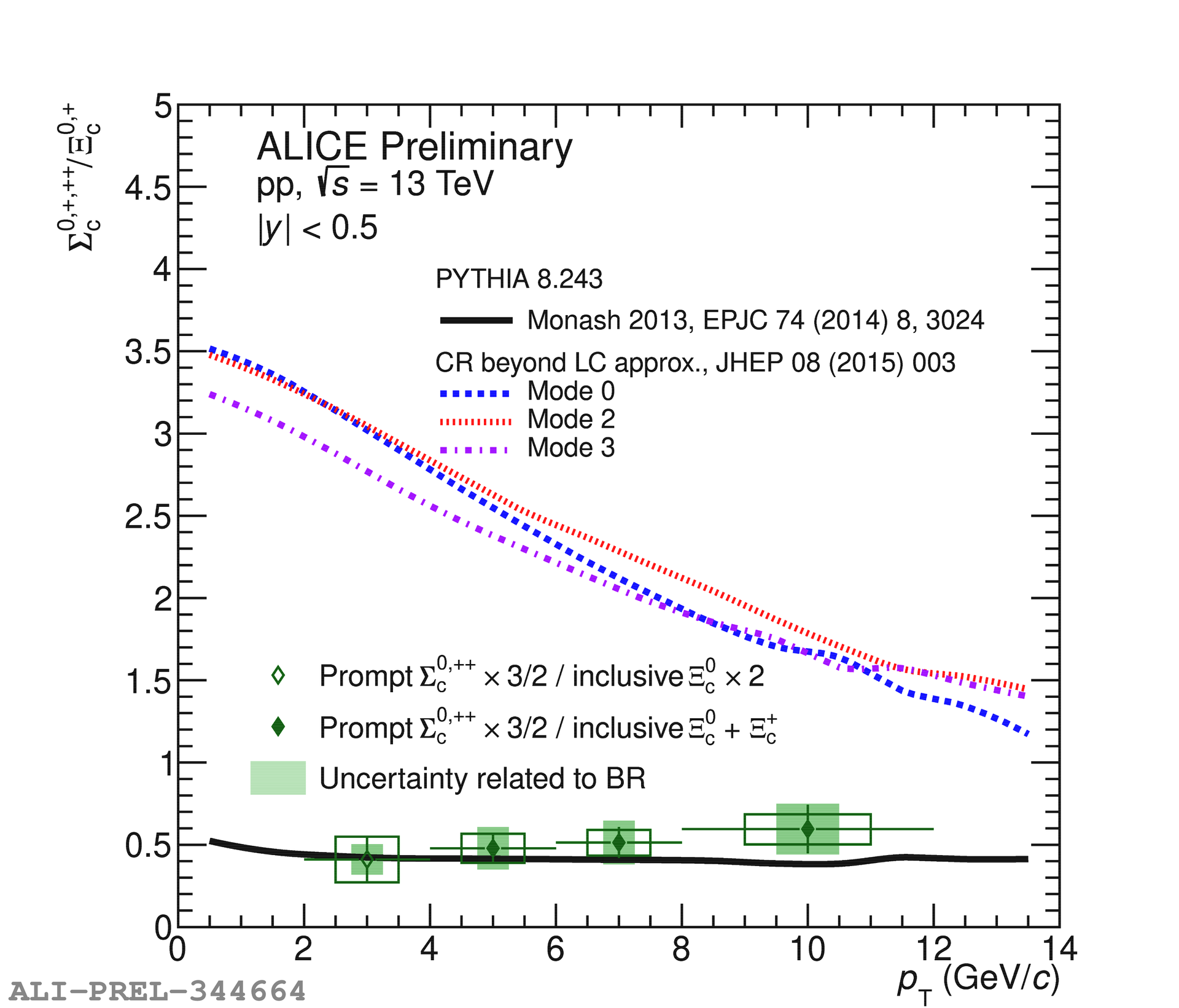}
\caption{\label{SoverX}Ratio of ${\rm \Sigma_c}$ and ${\rm \Xi_c}$ baryons compared with PYTHIA8 predictions with different tunes.}
\end{minipage} 
\end{figure}

In Figs.~\ref{SoverL} and \ref{SoverX} baryon-to-baryon ratios are shown. In the case of ${\rm \Sigma_c}$-to-${\rm \Lambda_c}$ ratio (Fig.~\ref{SoverL}) all the PYTHIA8 tunes \cite{Christiansen:2015yqa, Skands:2014pea} cannot describe the data. PYTHIA uses the fragmentation functions based on the ${\rm e^+e^-}$ collisions, which may signal that charm hadronization strongly depends on the collision system. The SHM-based model \cite{He:2019tik}, on the other hand, reproduces well the ratio of non-strange charmed baryons. For the ratio of ${\rm \Sigma_c}$ and ${\rm \Xi_c}$ (Fig.~\ref{SoverX}), the measurements are well described by the PYTHIA8 Monash tune. However, the higher-order colour string formation scenario predicts further relative enhancement in charmed-baryon production, which is not supported by data.

\section{Multiplicity-dependent production of heavy-flavour hadrons}

\begin{figure}[ht!]
    \centering
    \includegraphics[width=0.6\textwidth]{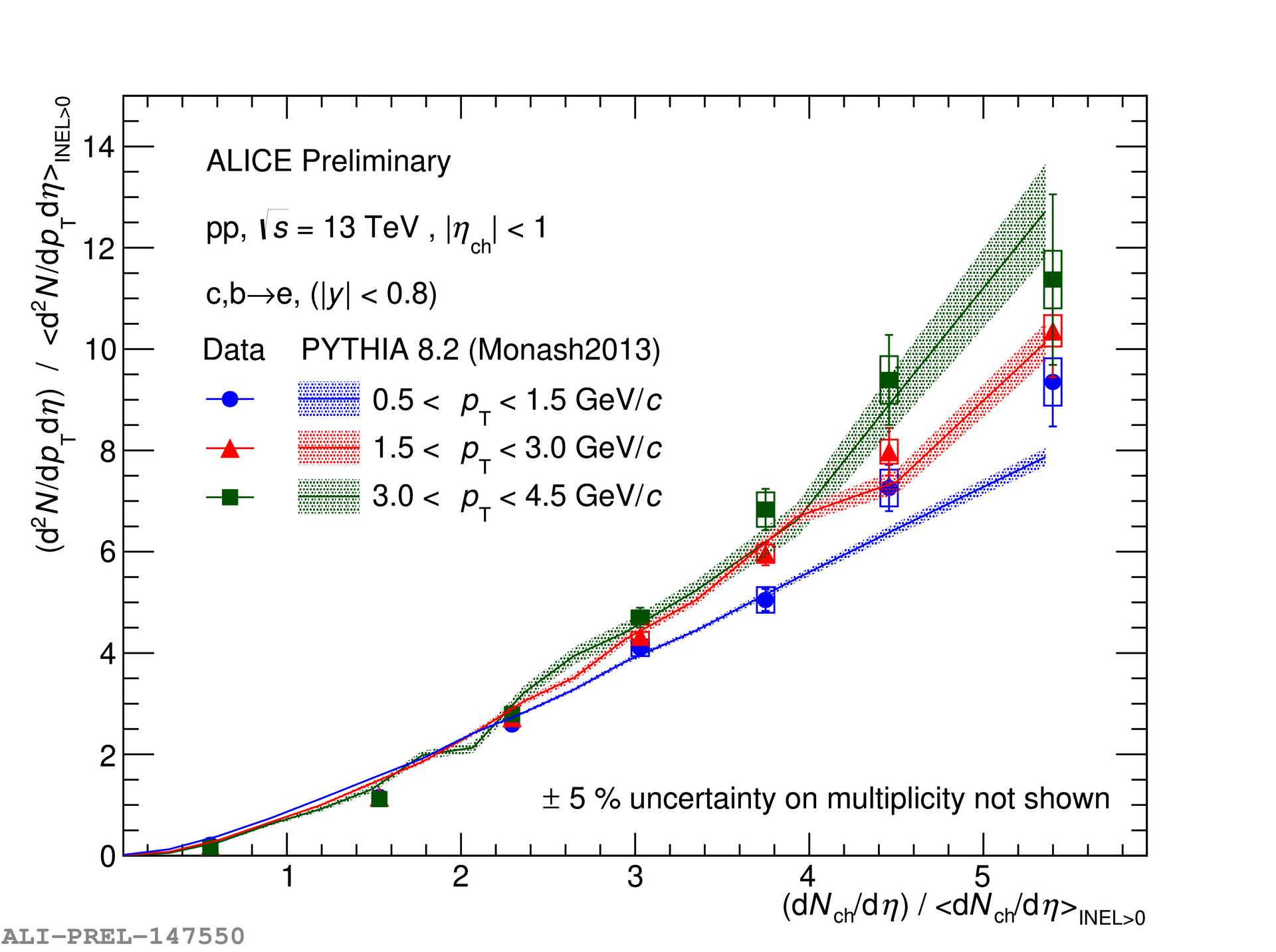}
    \caption{Self-normalised yields of electrons from heavy-flavour hadron decays as a function of self-normalised multiplicity measured at midrapidity in different $p_{\rm T}$ bins.}
    \label{multe}
\end{figure}

In multiplicity-dependent studies of particle production, it was observed that the relation between the self-normalised multiplicity and the self-normalised number of produced particles at midrapidity is stronger than linear. This can be partly explained by the presence of the autocorrelation effects arising from defining the multiplicity in the same region as the signals. Models including MPI effects \cite{Sjostrand:2014zea}, or similar mechanisms \cite{Drescher:2000ha,Ferreiro:2012fb}, also predict this kind of behaviour (see Fig.~\ref{multe}). This stronger-than-linear dependence is also observed for ${\rm J/\psi}$, B and D mesons \cite{Abelev:2012gx, Adam:2016mkz}.

\begin{figure}[ht!]
\begin{minipage}{0.45\textwidth}
\includegraphics[width=\textwidth]{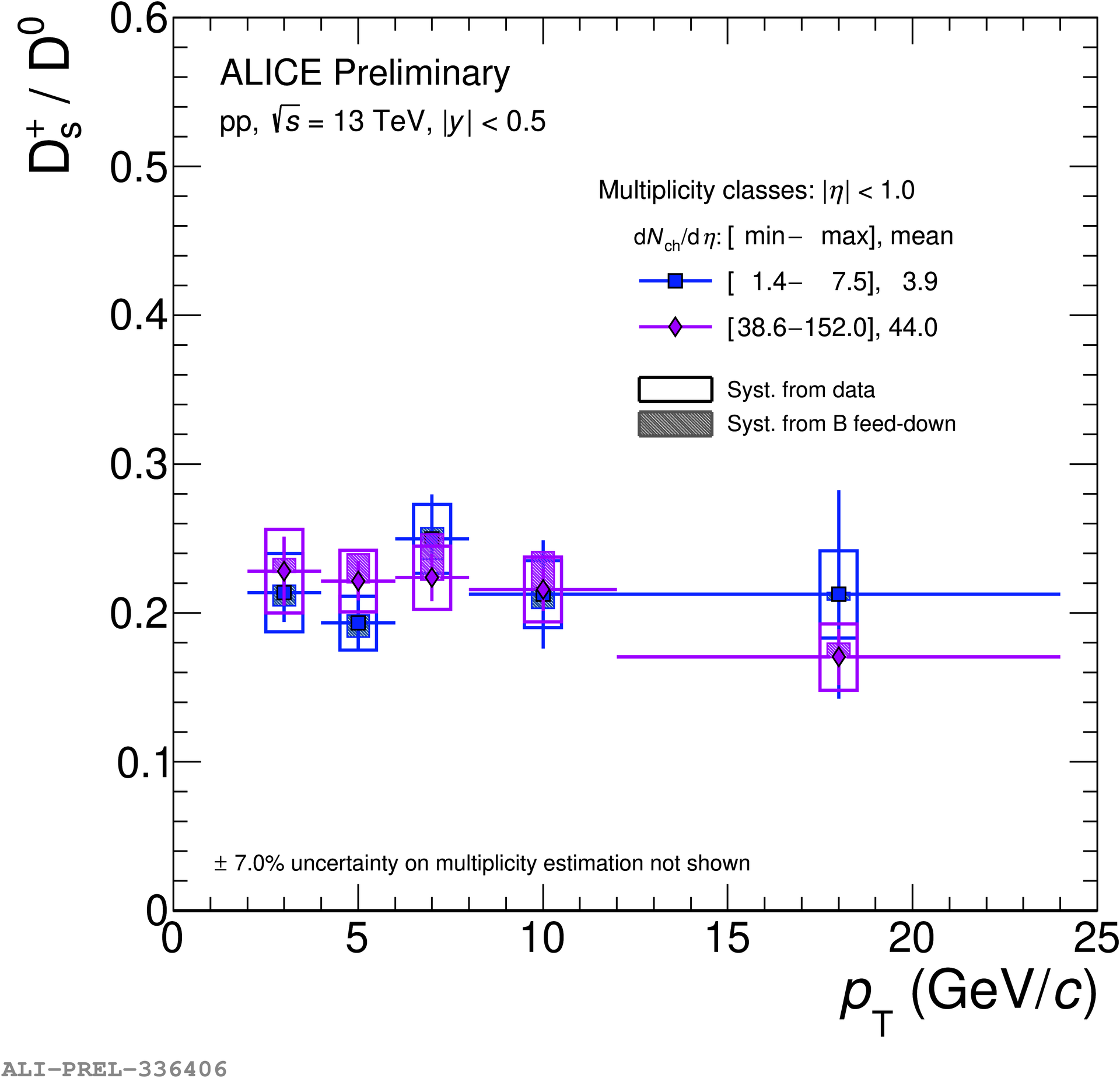}
\caption{\label{multS}Ratio of ${\rm D_s^+}$ and ${\rm D^0}$ mesons at low and high multiplicity.}
\end{minipage}\hspace{0.1\textwidth}%
\begin{minipage}{0.45\textwidth}
\includegraphics[width=\textwidth]{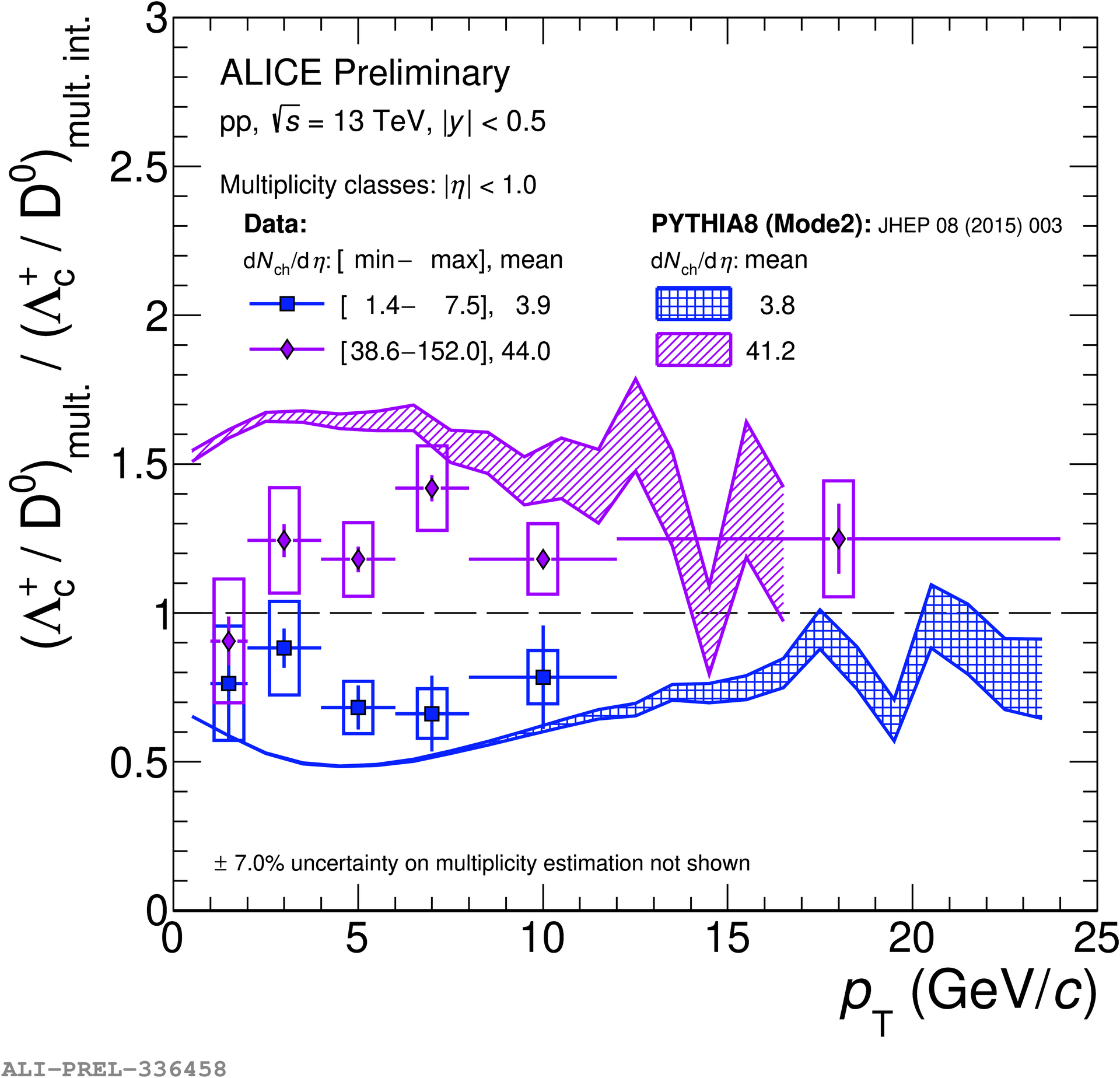}
\caption{\label{multL}Self-normalised ratio of ${\rm \Lambda_c^+}$ and ${\rm D^0}$ at low and high multiplicity compared with the PYTHIA8 Mode2 simulations.}
\end{minipage} 
\end{figure}

Strangeness enhancement has been found to be dependent on the event multiplicity, regardless of the collision system \cite{ALICE:2017jyt}, suggesting that the production of light flavour and strange hadrons is dominated by the final state. In order to study the influence of strangeness content on heavy-flavor production, a measurement of ${\rm D_s^+/D^0}$ was executed in ALICE as a function of multiplicity. Results in Fig.~\ref{multS} show that the strange-charmed ${\rm D_s^+}$ meson production in pp collisions is not influenced by the multiplicity. This is different from the behaviour measured for light-flavour particles. 

Figure \ref{multL} shows a clear distinction between the hadronization of charmed ${\rm \Lambda_c^+}$ baryons at low and high multiplicity. This behaviour is predicted by the PYTHIA8 with Mode2 tune, however description is only qualitative with this tune.

\section{Conclusions}

The ALICE collaboration performed a broad range of measurements on heavy-flavour particle production in pp collisions in order to test perturbative QCD models. The FONLL calculations predict well the production cross section of heavy-flavour particles, although the data are at the upper limit of the theoretical uncertainties. Also, measurements studying the hadronization of mesons and baryons are executed. PYTHIA8 Monash tune is insufficient to describe the hadronization processes, however the Mode2 tune with higher-order colour reconnection calculations performs better in this. Another model, based on  statistical hadronisation with augmented charmed baryon states, also provides qualitatively good predictions for the ratio of non-strange ${\rm \Lambda_c}$ and ${\rm \Sigma_c}$ baryons. However neither of these two models can describe the enhancement of ${\rm \Xi_c}$ baryons with respect to ${\rm D^0}$.

ALICE also performed multiplicity-dependent measurements in order to study the connection of underlying events to the leading hard process. The results show a stronger than linear dependence of heavy-flavour production on event multiplicity, that can be attributed to Multiple Parton Interactions. A multiplicity-dependent comparison of strange ${\rm D_s}$ and non-strange D mesons does not show an enhancement of strange D mesons relative to non-strange D mesons with multiplicity in pp collisions. On the other hand, a clear multiplicity dependence is observed for ${\rm \Lambda_c}$ baryons when compared to ${\rm D^0}$ mesons.

\section*{Acknowledgements}

This work has been supported by the Hungarian NKFIH/OTKA FK131979 and K135515 grants, as well as the NKFIH 2019-2.1.6-NEMZ\_KI-2019-00011 and 2019-2.1.11-TÉT-2019-00078 projects.

\section*{References}

\bibliography{refs}

\end{document}